\documentclass[structabstract]{aa} 
\usepackage{graphicx}
\usepackage{txfonts}
\usepackage{amsmath}
\usepackage{pslatex}
\usepackage{hyperref}
\usepackage[all]{hypcap}
\usepackage{cleveref}
\begin{document}

   \title{Relativistic effects on coronal ejection in variable X-ray sources}

   \author{B. Mishra
     \and W. Klu\'zniak
          }

   \institute{
     Nicolaus Copernicus Astronomical Center,
     Polish Academy of Sciences,
     ul. Bartycka 18, 00-716 Warszawa, Poland
     \\~e-mail:mbhupe@camk.edu.pl, wlodek@camk.edu.pl
                 }
   \date{Received ***; accepted ***}
  \abstract
     {{\it Context:} Optically thin coronae around neutron stars
     suffering an X-ray burst can be ejected as a result of  rapid
     increase in stellar luminosity. In general relativity (GR),
     radiation pressure from the central luminous star  counteracts
     gravitational attraction more strongly than in Newtonian physics.
     However, motion near the neutron star is very effectively impeded
     by the radiation field.\\
 {\it Results:} We discuss coronal ejection
     in a general relativistic calculation of the motion of a test
     particle in a spherically symmetric radiation field. At every
     radial distance  from the star larger than that of the ISCO, and
     any initial luminosity of the star, there exists a luminosity
     change which leads to coronal ejection. The luminosity required
     to eject from the system the inner parts of the optically thin
     neutron-star corona is very high in the presence of radiation
     drag and always close to the Eddington luminosity. Outer parts of
     the corona, at a distance of ~20 $R_G$ or more, will  be
     ejected by a sub-Eddington outburst. Mildly fluctuating
     luminosity will lead to dissipation in the plasma and may explain
     the observed X-ray temperatures of coronae in low mass X-ray
     binaries (LMXBs).  At large radial distances from the star
     ($3\cdot 10^3 R_G$ or more) the results do not depend on whether or not
     Poynting-Robertson drag is included in the calculation.}
  


   \maketitle
   
\section{Introduction}
Most LMXBs vary in luminosity on many
timescales.  Particularly rapid and luminous variations are exhibited by the
numerous X-ray bursters, which are thought to be neutron stars
undergoing a thermonuclear explosion on their surface, yielding an
Eddington luminosity at maximum.  Sometimes the maximum flux corresponds to
super-Eddington luminosities, as in the pulsating neutron star
 ``LMC transient'' A0535-668
(at a firm distance of 50 Kpc), which is thought to attain an
isotropic flux  $L_\infty=1.2\times10^{39}{\rm erg/s}=6.9\,L_{\rm Edd}$
for a $1.4\, M_\odot$ star \citep{Bradt}. One may ask whether the variability of
X-ray luminosity has any influence on the state of circumstellar matter.

 The observed LMXBs are thought to be powered by accretion occurring through an optically thick disk. 
In this paper we are considering the response
of accretion flows in the optically thin regions of the system
 to X-ray variability of the source.
 We are considering a corona in a variable X-ray source. In this context, corona means optically thin plasma ``above the surface'' of the disk. 
The effects of luminosity change on the motion of optically thin accreting matter in Newtonian dynamics are  already known, an impulsive increase in the stellar luminosity of the star may lead to coronal ejection if the luminosity is increased by one half or more of the difference between Eddington luminosity and the initial one \citep{2013A&A...551A..70K}. However, that result was derived analytically taking no account of radiation drag.
\href{ram}{\citet{1989ApJ...346..844W}} pointed out the effect of Poynting-Robertson drag \citep{1937MNRAS..97..423R} for accretion flows around high luminosity neutron stars. Proper inclusion of radiation drag requires a numerical solution in general relativity (GR), and we are presenting such solutions in this paper.
Recently, \cite{stahl13}
 discussed the problem for super-Eddington outbursts in initially non-radiating stars. Here, we consider a wide range of initial and final luminosities.

We are attempting to understand luminosity effects on the corona by modeling test-particle motion around the star. The dynamics are described by equations of motion of a particle moving in a spherically symmetric radiation field in the Schwarzschild metric, while interaction with the radiation by momentum absorption with a cross-section whose numerical value will correspond to the Thomson cross-section times the mass of the particle expressed in units of proton mass. This assures that the conventional Eddington luminosity, $L_{\rm Edd}$, will balance gravity exactly for hydrogen plasma at infinity (i.e., in the Newtonian limit). The calculation can be carried over to other compositions, and other cross-sections for photon absorption, by suitably redefining the Eddington luminosity.

The radiation, assumed to be originating on the surface of a neutron star, will have two important effects on the motion: first, it will counteract gravity by transfering the radial component of (a part of) its momentum; second, it will exert a drag on the moving particle. The calculation is performed in GR, and includes all non-vanishing components of the radiation stress tensor.
 
\par
The paper is organized in the following manner. In \S~\ref{eom} we shall briefly explain the equations of motion that describe the trajectories of test particles. In \S~\ref{ssnodrag}, in order to isolate the gravitational effects of GR, we shall present results of a simplified calculation without radiation drag, and in \S~\ref{ssdrag} we will describe the test particle behavior determined from the complete equations of motion (e.o.m.), which include radiation drag. In \S~\ref{conclude}  we  discuss the results and present the conclusions in \S~\ref{conclusion} . 

\section{Equations of motion}
\label{eom}
We performed all the calculations in the Schwarzschild metric, using spherical polar coordinates $(r,\theta,\phi)$. For a spherically symmetric radiation field, motion of a test-particle is restricted to one plane and we choose it to be  the  equatorial plane ($\theta= \pi/2$). \citet{1990ApJ...361..470A} performed a rigorous analysis of purely radial motion of a test particle in the combined gravity and isotropic radiation fields of a spherical, non-rotating, compact star, and we shall use their stress energy tensor of radiation, $T^{(\mu)(\nu)}$, calculated in a stationary observer's tetrad assuming isotropic emission from the surface of star, see also \cite{stahl13}. Numerical  solutions of the equations of motion of test particle trajectories in a steady radiation field have been obtained e.g., by \citet{2009CQGra..26e5009B} and \citet{2012A&A...546A..54S} in the Schwarzschild metric, and by \citet{2010PhRvD..81h4005O,Semerak} in the Kerr metric.

 To describe in detail the effects of {\it variable luminosity} on the motion of optically thin plasma, and specifically the effects of impulsive changes in luminosity on test-particle trajectories, we use two sets of equations in the present paper. One, a simplified set, describes the trajectories of test particle without considering radiation drag, and the other includes all components of the radiation stress tensor, including the terms responsible for radiation drag. 

\subsection{Complete equations of motion (``with drag'')}
We performed the computations using dimensionless co-ordinates scaled by gravitational radius $R_G$, i.e., the dimensionless radial position, $x$, stellar radius, $X$, and proper time, $\tau$,
$$
d\tau=\frac{\rm d s}{R_G},
\ x=\frac{r}{R_G},
\ X=\frac{R}{R_G},
$$
where the gravitational radius is 
$
R_G= {GM}/{c^2}.
$
It is convenient to abbreviate the metric coefficient as $B=(1-2/x)$.
The e.o.m. are \citep[e.g.,][]{2012A&A...546A..54S,stahl13}
\begin{equation}
\frac{d^2 x}{d \tau^{2}}= \frac{k}{\pi I(R)X^2}\left(B T^{(r)(t)}u^{t}- \left[T^{(r)(t)}+\epsilon\right]\frac{d x}{d \tau}\right)
 +  
\label{radialfull}
\end{equation}

\begin{center}
$$
+ \left(x-3\right)\left(\frac{d \phi}{d \tau}\right)^2-\frac{1}{x^2},
$$
\end{center}

\begin{equation}
\frac{d^2 \phi}{d \tau}=-\frac{d \phi}{d \tau}\left(\frac{k}{\pi I(R)X^2}\left[T^{(\phi)(\phi)}+
\epsilon\right]+\frac{2}{x}\frac{d x}{d \tau}\right),
\label{azimuthfull}
\end{equation}
where
\begin{eqnarray} 
\epsilon \!&=&\!
BT^{(t)(t)}(u^t)^2+B^{-1}T^{(r)(r)}\left(\frac{d x}{d \tau}\right)^2 +\\
&& +  x^2T^{(\phi)(\phi)}\left(\frac{d \phi}{d \tau}\right)^2 
- 2T^{(r)(t)}u^t\frac{d x}{d \tau},
\label{varepsilon}
\end{eqnarray}
and the time component of the particle four velocity is given by
\begin{equation}
u^t =B^{-\frac{1}{2}}\left[1+B^{-1}\left(\frac{d x}{d \tau}\right)^2 +x^2\left(\frac{d \phi}{d \tau}\right)^2\right]^{\frac{1}{2}}.
\end{equation}

In all the computations the luminosity  at the surface of the star,  $L(R)$, is represented in units of Eddington luminosity as $k$ (Eqs.~\ref{radialfull}, \ref{azimuthfull}). One can also display the results in terms of the luminosity at infinity, $L_{\infty}$. These are related through
\begin{equation}
k=\frac{L(R)}{L_{\rm Edd}}=\frac{L_{\infty}}{L_{\rm Edd}}\left(1-\frac{2}{R}\right)^{-1}.
\end{equation}

To obtain $x(\tau)$ and $\phi(\tau)$ we integrated  Eq.~\ref{radialfull} and Eq.~\ref{azimuthfull} using the code of
\cite{2012A&A...546A..54S} and \cite{2012A&A...545A.123W}, relying on the Dormand-Prince method, which is a fourth-order accuracy, adaptive step-size Runge-Kutta type integration method. In all our simulations we assumed a fixed radius of neutron star of $X=5$. At this stellar radius, the marginally stable radius (ISCO) is outside the surface of the neutron star \citep{kw}.

\begin{figure}
\centering
\includegraphics[width=5cm,bb = 130 50 581 570]{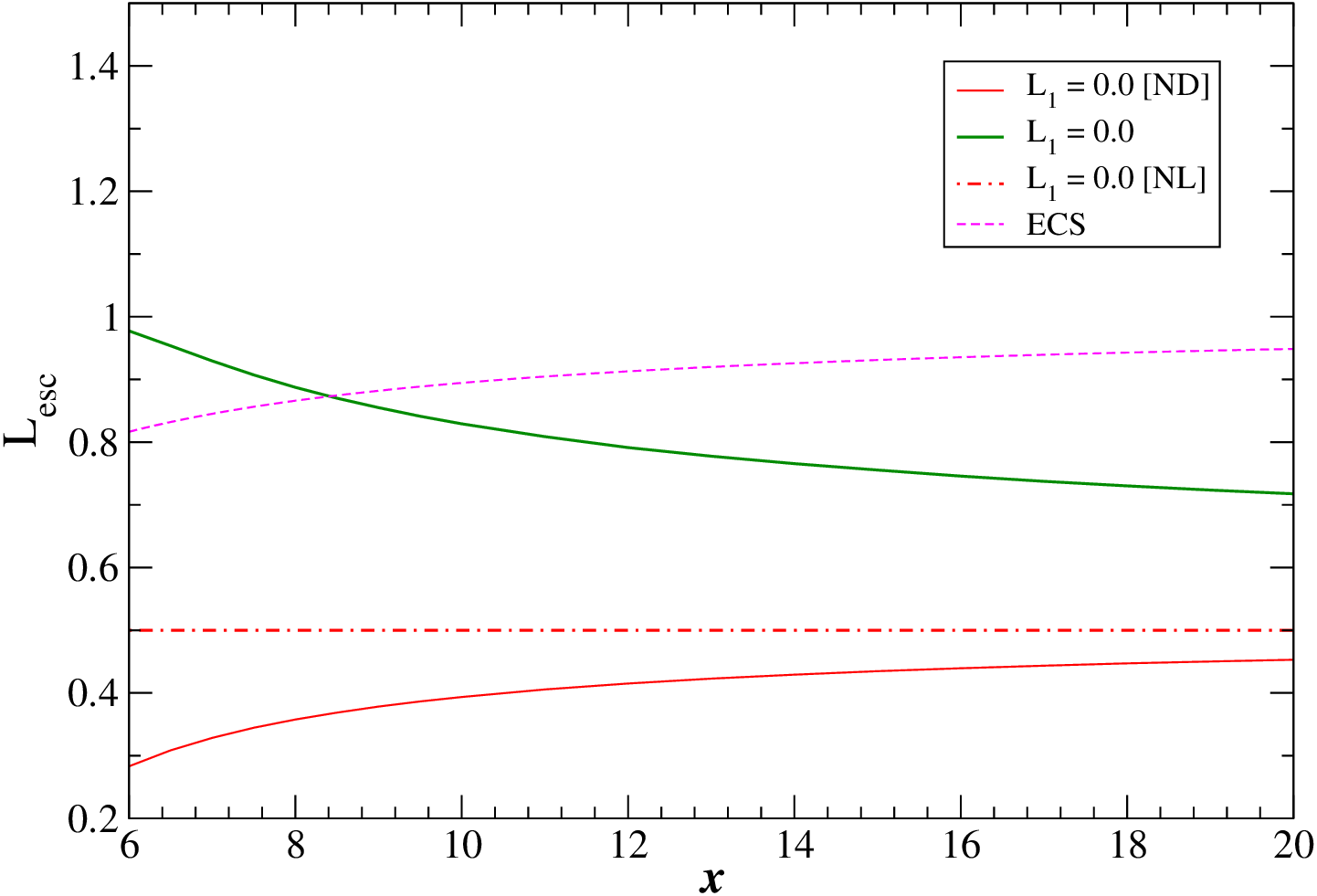}
\caption{Minimum luminosity that will eject to infinity a particle initially in circular orbit at $x$. Plot compares solutions of the full e.o.m. (solid green curve, Eqs.~\ref{radialfull}, \ref{azimuthfull}) with solutions of the simplified equations that neglect the effects of radiation drag (``[ND]'' thin red line, Eqs.~\ref{radial}, \ref{azimuth}). Initial luminosity is $L_1=0$, stellar radius $X=5$. The Newtonian limit is also shown (dot-dashed line), as well as the radius of the Eddington capture sphere (ECS). See text for details.
\bigskip}
\label{comparison}
\end{figure}

\begin{figure}
\centering
\includegraphics[width=5cm,bb = 130 50 581 532]{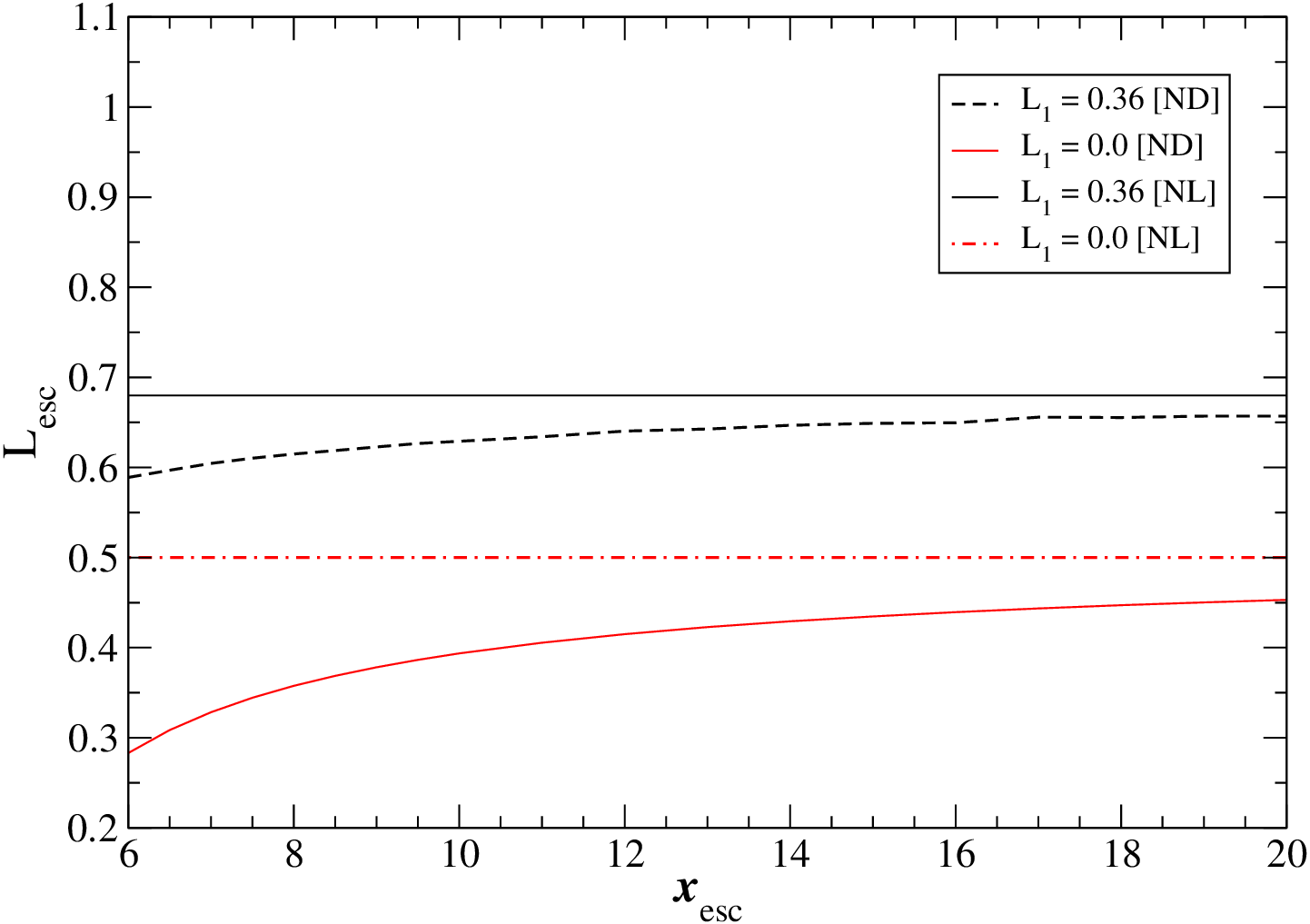}
\caption{Radius of the escape sphere  when drag is neglected. The minimum luminosity, $L_{\rm esc}$, that unbinds a particle in circular orbit at $x_{\rm esc}$ is shown as a function of the initial orbital radius, for two values of initial luminosity, $L_1=0.0$, and $0.36$. Stellar radius is $X=5.0$. Constant $L_{\rm esc}$ lines represent Newtonian limits (NL) for $L_1=0.0,0.36$. 
If there were no drag, particles from all circular orbits of radii in the range $6\le x(0)<x_{\rm esc}(L_{\rm esc})$  would escape the system, and the initial radius in the outgoing trajectory, $x(0)$, would be its periastron as well.
}
\label{nodrag}
\end{figure}

\begin{figure}
\centering
\includegraphics[width=5cm,bb = 130 45 581 570]{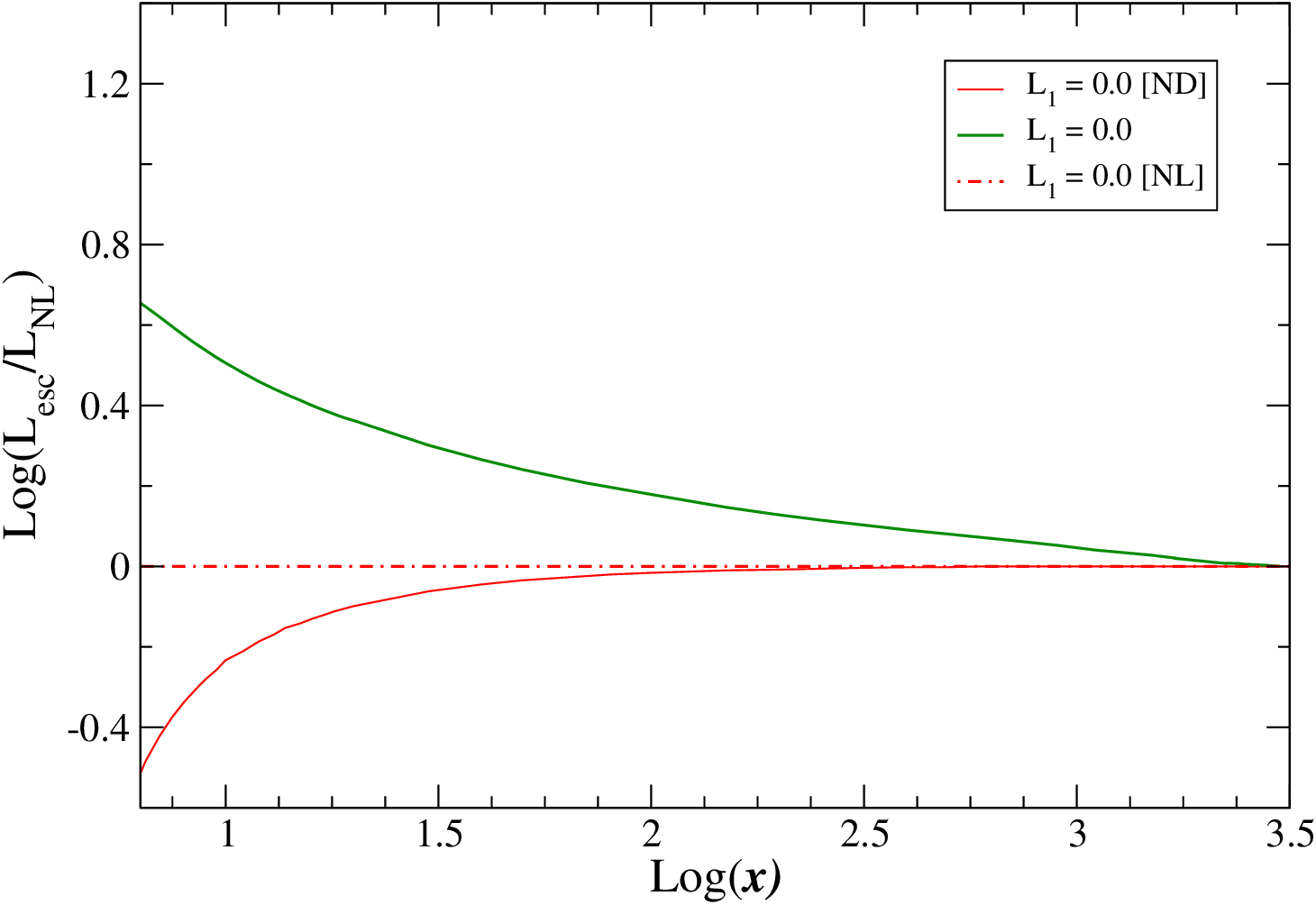}


\caption{Plot shows convergence of results using Eqs. 1, 2 and Eqs. 6, 7, with the Newtonian limit. The ratio is plotted of the minimum luminosity sufficient to unbind a particle, initially in a circular GR orbit at $L_1 = 0$, to the corresponding value in the Newtonian limit.
Uppermost (thick green) curve corresponds to Eqs.~\ref{radialfull} and \ref{azimuthfull},  and red continuous curve to Eqs.~\ref{radial} and \ref{azimuth}.}
\label{newtonlim}
\end{figure}

\subsection{Equations of motion without drag}

Neglecting the terms responsible for radiation drag, one obtains simplified equations of motion involving only one component of the radiation stress tensor,

\begin{equation}
\frac{d^2 x}{d \tau^{2}}= \frac{k}{\pi I(R)X^2}B T^{(r)(t)}u^{t}
+ \left(x-3\right)\left(\frac{d \phi}{d \tau}\right)^2-\frac{1}{x^2},
\label{radial}
\end{equation}
\begin{equation}
\frac{d^2 \phi}{d \tau}=-\frac{d \phi}{d \tau}\left(\frac{2}{x}\frac{d x}{d \tau}\right).
\label{azimuth}
\end{equation}

\section{Coronal ejection}
\label{ejection}
If a particle is moving in a circular orbit, it can become unbound under a sudden increase of luminosity. For a test particle in Newtonian dynamics, moving initially in a Keplerian circular orbit under initial luminosity $L_1$, the minimum luminosity change required to unbind the particle is $(L_{\rm Edd}-L_1)/2$.  Such a luminosity change will also lead to coronal ejection if the radiation drag is neglected \citet{2013A&A...551A..70K}. 

We will be contrasting conditions under which particles escape to infinity with those in which the particles remain in the system. The results presented below are based on numerical solutions for the trajectories. Therefore, we need a numerical criterion for deciding which outgoing trajectories are bounded, and which ones are unbounded. We adopted one\footnote{This criterion differs in some respects from the one adopted by \cite{stahl13}.} in which we computed the Newtonian  effective specific energy of the test particle at $x=3000$. By the effective specific energy we mean the specific energy in the potential $-GM_{\rm eff}/r$,
 with $M_{\rm eff}=M(1-L_{\infty}/L_{\rm Edd})$; this prescription takes account of the radial radiation pressure term.
If the Newtonian effective specific energy is positive at $x=3000$, the trajectory is deemed to be unbounded (i.e., extending to infinity), and if it is negative, the trajectory is deemed bounded. In practice, at such a large distance the trajectory of those particles which have reached it is always very nearly radial, and the influence of drag is negligible there.

To isolate the effects of drag from gravitational effects in GR
we shall discuss two cases, in the first case we shall omit effects of radiation drag (Eqs.~\ref{radial}, \ref{azimuth}), and in the second case we shall use the complete e.o.m. (Eqs.~\ref{radialfull}, \ref{azimuthfull}) that include also the radiation drag terms. 
Our results show convergence of the two cases at large radii
 (Fig.~\ref{newtonlim}).
Interestingly, radiation drag dominates GR corrections, and already at a radius
of about ~$10^2R_G$ the non-drag GR corrections appear to be negligible.
At $r=10^3R_G$ the fully relativistic solution (including drag) differs from the Newtonian limit by just a few per cent.

In the following we  give the value of luminosities at infinity scaled by
the Eddington luminosity. Thus, a ``luminosity of $L_1$'' signifies
 $L_{\infty}=L_1L_{\rm Edd}$.

\begin{figure}
\centering
\includegraphics[width=5cm,bb = 130 50 581 570]{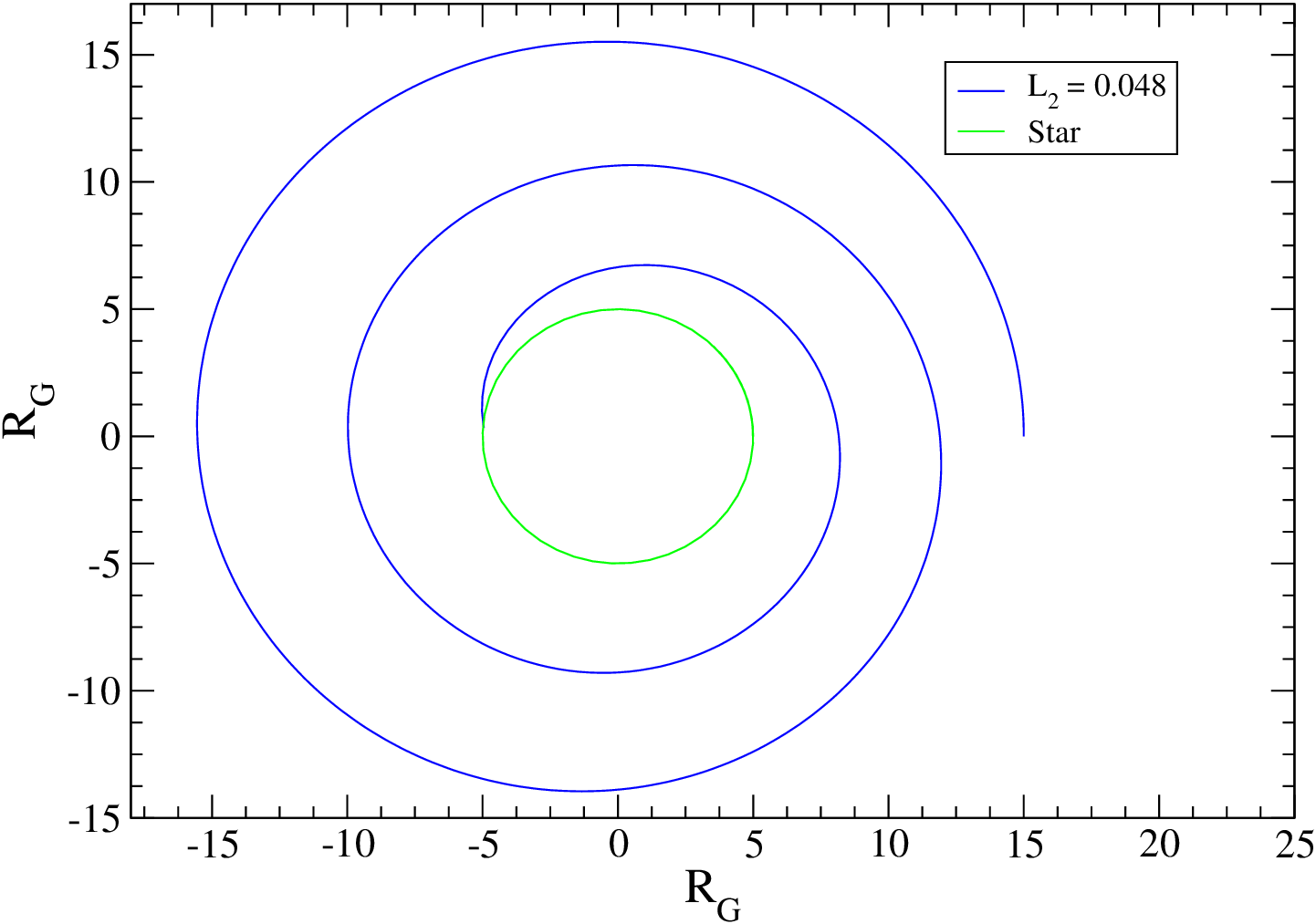}
\caption{Plot shows test-particle trajectory for a constant $L = 0.048$. Instead of maintaining a circular orbit the test particle follows a spiral trajectory and accretes on the surface of star.}
\label{spiral}
\end{figure}


\subsection{GR effects without drag}
\label{ssnodrag}
We consider an initial luminosity, $L_1$ which at some point in the evolution of the system will be changed impulsively to a different value $L_2$.
In this subsection we assume the absence of radiation drag. Thus a test particle can move in a Keplerian orbit at a fixed radial distance from the star, as long as the luminosity is held fixed. It will continue to move on nearby orbits under very small changes in the luminosity. But if the luminosity change is large enough, the particle can escape the system.

We investigate for what luminosity change $L_2-L_1$ the particle will become unbound (neglecting radiation drag).
The initial conditions correspond to a circular orbit, at a particular radius around the star of luminosity $L_1$. To find the trajectory we integrate
Eqs.~\ref{radial}, \ref{azimuth}.
Since the effective radiative force diminishes more rapidly  with increasing distance from the center of a variable X-ray source than the gravitational attraction, in the absence of drag a smaller luminosity change is required to eject the particle from a circular orbit close to the ISCO ($x=6$ for a non-rotating neutron star) than from a more distant one. This is clear in Fig.~\ref{nodrag}. 
The plotted value, $L_{\rm esc}(x_{\rm esc})$, corresponds to the minimum luminosity change $L_{\rm esc}-L_1$ sufficient to unbind a particle that was in a circular orbit at $x=x_{\rm esc}$ at luminosity $L_1$, assuming that drag is neglected.
Also shown (constant $L_{\rm esc}$ lines in the figure)
 is the corresponding Newtonian limit of
 \citet{2013A&A...551A..70K}.

 It turns out that for any given pair of initial and final luminosities there is a sphere, inside of which the luminosity change is sufficiently large for all orbits to become unbounded---all particles within the sphere abandon their previously stable circular orbits for trajectories extending to infinity. On the other hand, particles previously orbiting outside the sphere remain bound. 
We will denote the radius of this {\it escape sphere} by $x_{\rm esc}$.
Another way of looking at Fig.~\ref{nodrag} is that it presents the escape sphere radii (in motion without drag) as
a function of the final luminosity $L_{\rm esc}$, for two different initial luminosities.

\begin{figure}
\centering
\includegraphics[width=5cm,bb = 130 50 581 570]{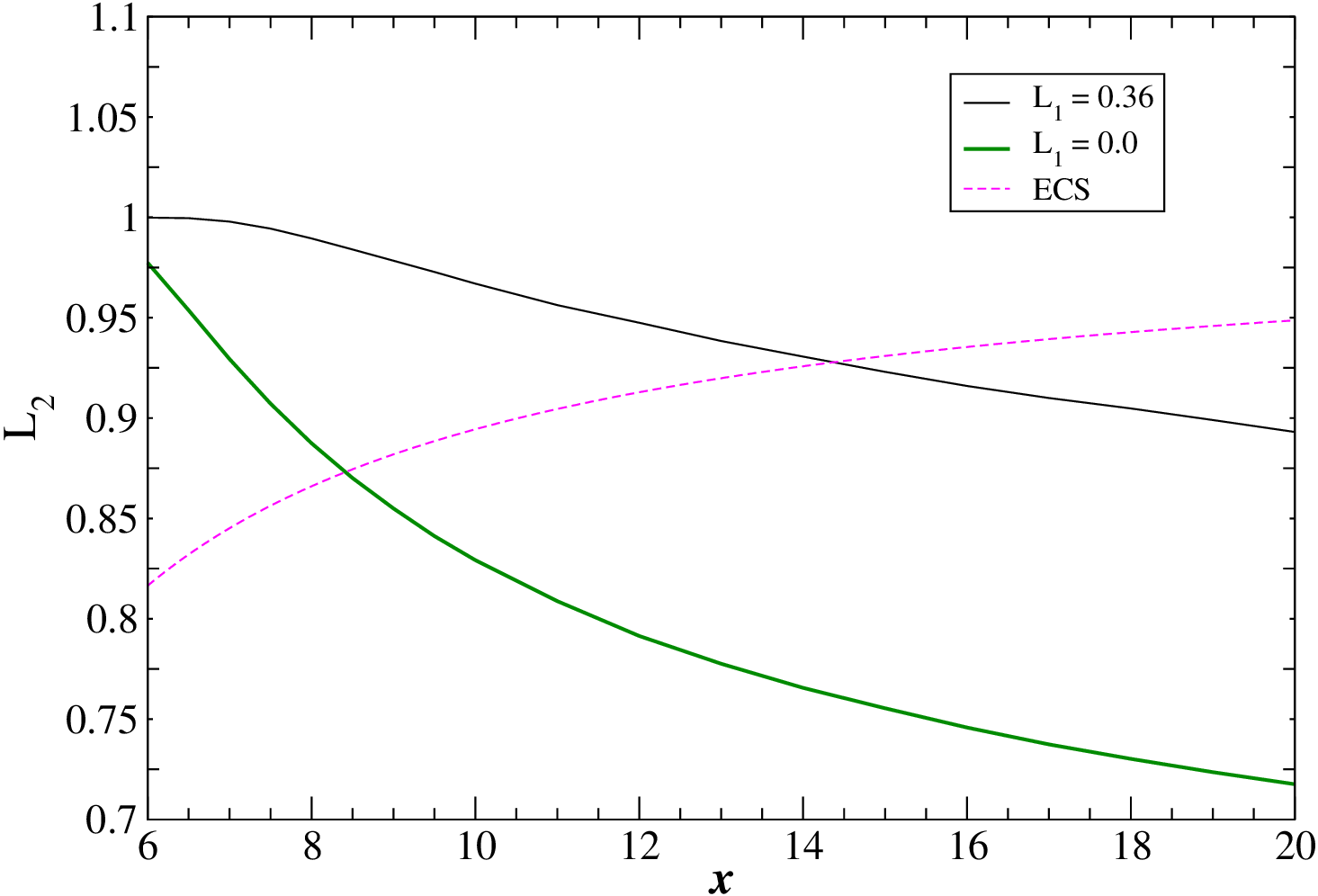}
\caption{The continuous curves show the variation with the final luminosity of minimum escape radius, $x=x_{\rm esc}(L_2)$, after a sudden change to $L_2$ from $L_1=0$, or $0.36$. The same curves also show the minimum final luminosity, $L_2=L_{\rm esc}(x)$ required to eject to infinity a particle previously orbiting at $x$.
Stellar radius is $X=5.0$. Dotted curve represents the variation of ECS radius with $L_2$ (see text).}
\label{drag}
\end{figure}
\subsection{GR effects with drag}
\label{ssdrag}
Now we shall integrate Eq.~\ref{radialfull} and Eq.~\ref{azimuthfull} to compute the trajectories of test particles suffering the influence of the full radiation tensor. Radiation drag is automatically included in the formalism, and the angular momentum and energy of the test particle are no longer constants of motion---in this case we cannot get an analytical expression like in Newtonian dynamics by integrating the equations. In a steady radiation field, a circular orbit cannot be maintained---drag will cause the orbiting test particle to loose angular momentum and energy, and will cause it to follow a spiral trajectory \citep{1989ApJ...346..844W}, cf. Fig.~\ref{spiral}.

If initially a test particle is moving  under the influence of luminosity $L_1$, a sudden change to a different luminosity $L_2$ will change its trajectory. 
For an initially non-luminous star ($L_1=0$) the discussion is straightforward.
The initial trajectory may be taken to correspond to a stable circular orbit,
and the parameter $k$ in Eqs.~\ref{radialfull} and \ref{azimuthfull}
and may be taken to be non-zero starting at any moment. Again, for any radius of the initial circular orbit,
there is a minimum luminosity  $L_2=L_{\rm esc}$ which makes the test-particle unbound
(Fig.~\ref{comparison}). Note that now $L_{\rm esc}$ is a monotonically decreasing function of $x$. For comparison, also plotted in Fig.~\ref{comparison} are $L_{\rm esc}$ in the Newtonian limit \citep[dot-dashed curve,][]{2013A&A...551A..70K}, and in the GR calculation with radiation drag neglected (Eqs.~\ref{radial}, \ref{azimuth} and \S~\ref{ssnodrag}). Also shown (magenta dashed curve) is the effective Eddington luminosity, $L_{\rm eff}(x)$, i.e., that luminosity\footnote{$L_{\rm eff}(x)=(1-2/x)^{1/2}$.}
 at which  a test particle located at $x$ may be in a static equilibrium \citep[][]{Phinney}. The inverse function $x(L_{\rm eff})$
gives the radius of the Eddington capture sphere \citep{2012A&A...546A..54S,stahl13}, discussed below.

In the general case ($L_1\ne0$), the choice of initial conditions is less obvious.
Since we are interested in the change of the motion of accretion disk corona,
 resulting from the change in luminosity, we consider initialy circular orbits.
The initial condition for our calculations corresponds to the
motion of a test particle instantaneously in a circular orbit appropriate for stellar luminosity  $L_1$. Effectively, we assume that in a steady luminosity field the system will adjust to the presence of radiation drag, resulting in nearly circular motion of the accreting fluid.

Imagine a particle executing a  circular orbit around the star when it shines with luminosity $L_1$. This would only be possible if some external agency (e.g., the accretion disk) continually replenished the angular momentum and energy lost to radiation drag. The star is assumed to impulsively change its luminosity to $L_2$. We imagine the external agency to cease operation precisely at that instant when the spherical wavefront corresponding to the change in luminosity arrives at the orbit (for instance, the coupling between the disc and the corona is disrupted by the additional radiation pressure). Future hydrodynamic calculations will no doubt elucidate the true state of affair---ours is an exploratory calculation aiming to qualitatively  identify the main features in the response of a corona to variable illumination by the central star.
\begin{figure}
\centering
\includegraphics[width=5cm,bb = 130 50 581 570]{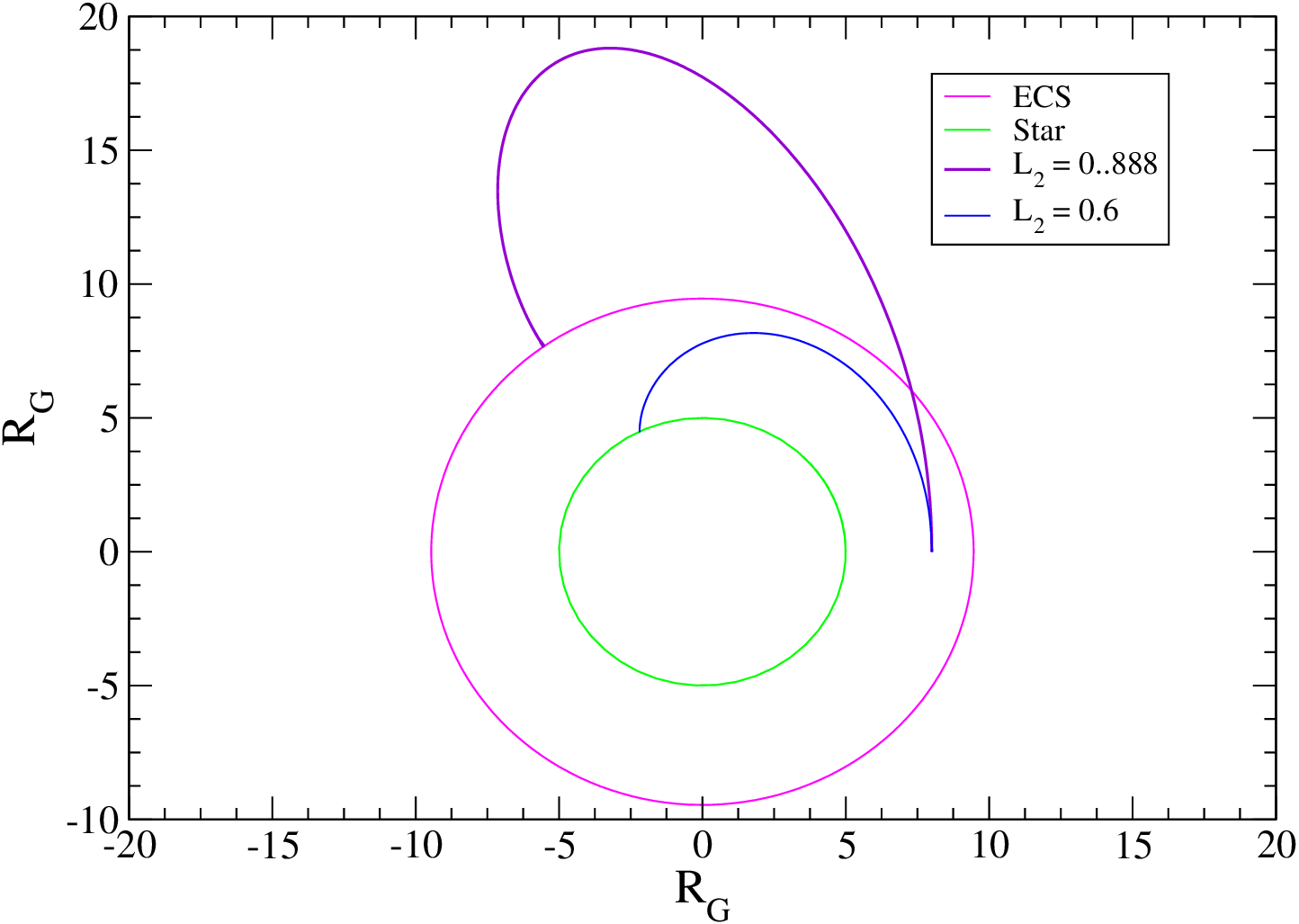}
\caption{Two qualitatively different trajectories with the same initial condition for two different luminosities: $L_2 = 0.6$ (thin blue trajectory) and $L_2 = 0.888$ (violet trajectory). For $L_2 = 0.888$ an ECS is present at $x_{\rm esc}= 9.44$. The radius of star $X=5.0$. The initial conditions are taken for $x=8.0$ and $L_1 = 0.36$ (cf. Fig.~\ref{drag}.) }
\label{outcome}
\end{figure}

Technically, we integrated Eq.~\ref{radialfull} and Eq.~\ref{azimuthfull} with a value of $k$ corresponding to the final luminosity, with the initial condition corresponding to a radius and velocity of a circular-orbit solution of  Eqs.~\ref{radial} and \ref{azimuth}
at luminosity $L_1$.  
The particle then follows a trajectory defined by its initial velocity and position, as well as the three stellar parameters $M$, $R$, $L_2$.

Again, we find that at any particular radius, $x$, there is a minimum luminosity $L_2=L_{\rm esc}$ that will cause a particle satisfying the initial conditions to escape to infinity. A final luminosity value larger than  $L_{\rm esc}$ will also lead to the expulsion of the particle.
The $L_{\rm esc}(x)$ curve is monotonically decreasing. 
Fig.~\ref{drag} illustrates this behavior.  The inverse function, $x_{\rm esc}(L_2)=x(L_{\rm esc})$  is the minimum radius of circular orbits from which the particles escape ({\it minimum escape radius}), for a given value of  $L_2$.
For a given value of  $L_2$, the sphere at $x_{\rm esc}$ now divides space into an outer region, $x\ge x_{\rm esc}$, from which the particles escape, and an inner region, $x< x_{\rm esc}$, the particles from which remain bound to the system (although they may leave the region  $x< x_{\rm esc}$).

Clearly, radiation drag is responsible for capturing all particles that are initially within
the sphere of escape radius $x_{\rm esc}$. At lower luminosities, the particles will actually be accreted onto the central star. However, at about Eddington luminosities they will lose all their momentum before settling on the star, and will levitate above the surface of the star in a state of equilibrium on a spherical surface concentric with the star, the so called {\it Eddington capture sphere} (ECS) \citep{2012A&A...546A..54S,stahl13}. 
The ECS exists at $X\le x<\infty$ for luminosity $L$ when $L_{\rm eff}(x)=L$
(cf. footnotes 2 and 3).

This is illustrated in Fig.~\ref{outcome}, where 
two trajectories are shown for motion starting with the same initial condition (instantaneously circular orbit at $x=8$ under luminosity $L_1=0.36$), but occurring for two different values of luminosities $L_2$, both satisfying $L_2<L_{\rm esc}(x)$, so that both motions are bounded. The two trajectories differ qualitatively in their termination point: at final luminosity  $L_2=0.6$ the particle is accreted with nonzero velocity onto the surface of the star, while at $L_2=0.888$ the ECS is present, on which the non-escaping particles come to rest. The latter case is an example of particles being trapped outside their initial position, radiation drag causing here a net displacement outwards, somewhat paradoxically.
The ECS luminosity-radius relation is shown in Fig.~\ref{drag}.
 The quasi-paradoxical net displacement outwards occurs only for radii to the left of the intersection of the ECS curve with the $L_{\rm esc}(x)$ curve, and only for values of $L_2$ between the two curves.
In Fig.~\ref{drag} the minimum escape radius has been presented for simulations with two values of initial luminosity, $L_1=0.0$ and $0.36$. We also plotted variation of ECS radius with the final luminosity, $L_2$. For a fixed initial luminosity $L_1$, the escape sphere and the ECS coincide at a particular radius (Stahl et al., 2013). In Fig.~\ref{drag} for initial luminosity $L_1=0.0$ and $0.36$, the minimum escape radius and the ECS radius are the same at $x_{\rm esc}=8.412$ and $ 14.365$, and corresponding luminosities $L_2=0.873$ and $0.928$, respectively.

\begin{figure}
\centering
\includegraphics[width=5cm,bb = 130 50 581 570]{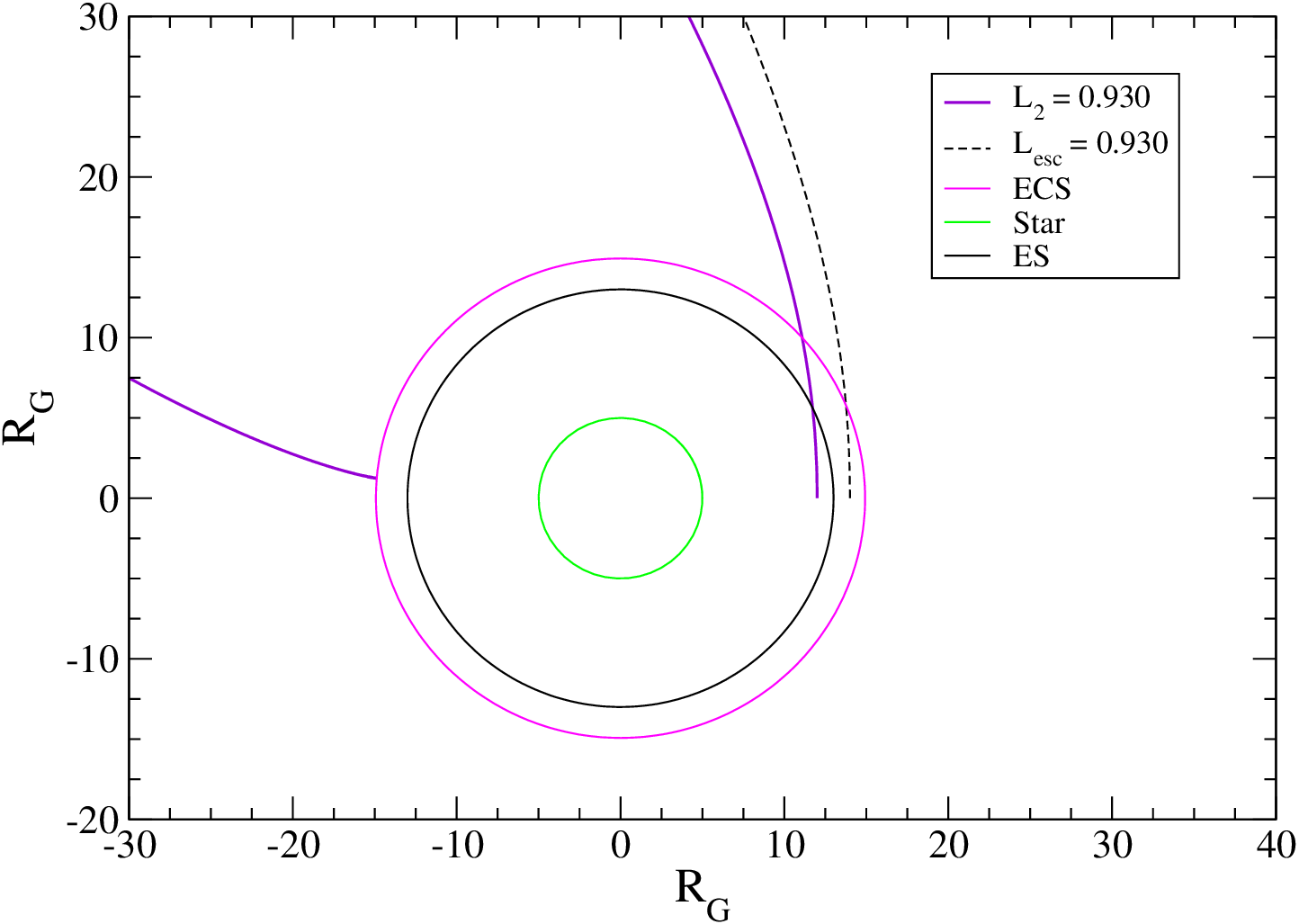}
\caption{Figure shows trajectories corresponding to the $L_1=0.36$ curve in Fig.~\ref{drag}. $L_2 = 0.930$ is fixed in both trajectories. The stellar surface is shown at $X=5.0$, and the ECS at $x_{ECS} = 14.927$ (magenta), the minimum escape radius is $x_{\rm esc}= 13.0$ (middle circle in black). Bounded (solid violet line) trajectory starts at $x=12.0$, unbounded (black dashed) trajectory starts at $x=14.0$. } 
\label{2tracks}
\end{figure}

Fig.~\ref{2tracks} is helpful in illustrating the concepts of the minimum escape radius $x_{\rm esc}$, and of the ECS. Here, we started with an initial luminosity of
$L_1=0.36$, impulsively changed to $L_2=0.930$. Two trajectories are illustrated
tangent to circles of radii $x_1=12.0$ and $x_2=14.0$. The initial velocities correspond to instantaneously circular motion under the influence of $L_1=0.36$, as described above. In this example the minimum escape radius is $x_{\rm esc}=13.0$, and clearly  $x_1<x_{\rm esc}$, while $x_2>x_{\rm esc}$. In the latter case [$x(0)=x_2$] the particle enters an escaping trajectory under the impulsive change of luminosity, while in the former [$x(0)=x_1$] it is captured by the ECS at $x=14.927$.

Two non-overlapping trajectories for the case of nonzero initial luminosity ($L_1 = 0.36$) are illustrated in Fig.~\ref{ECS}. The final luminosity is taken to be $L_2 =0.918$, and now the ECS is inside the sphere of minimum escape radius.
The black dashed trajectory illustrates an unbounded trajectory starting
form an instantaneously circular orbit of radius exceeding $x_{\rm esc}$.
The test particle on the blue continuous trajectory is captured on the ECS.
If the final luminosity is smaller than the one corresponding to an ECS at the stellar radius (at $X=5$), the non-escaping particles (i.e., ones with
 $x(0)<x_{\rm esc}$) will accrete on the star.\footnote{The ECS radius ranges from R to infinity for luminosity in the range $\sqrt{1-2R_G/R}\le L_\infty/L_{\rm Edd}<1$ \citep{stahl13}.}

\section{Discussion}
\label{conclude}
\begin{figure}
\centering
\includegraphics[width=5cm,bb = 130 50 581 568]{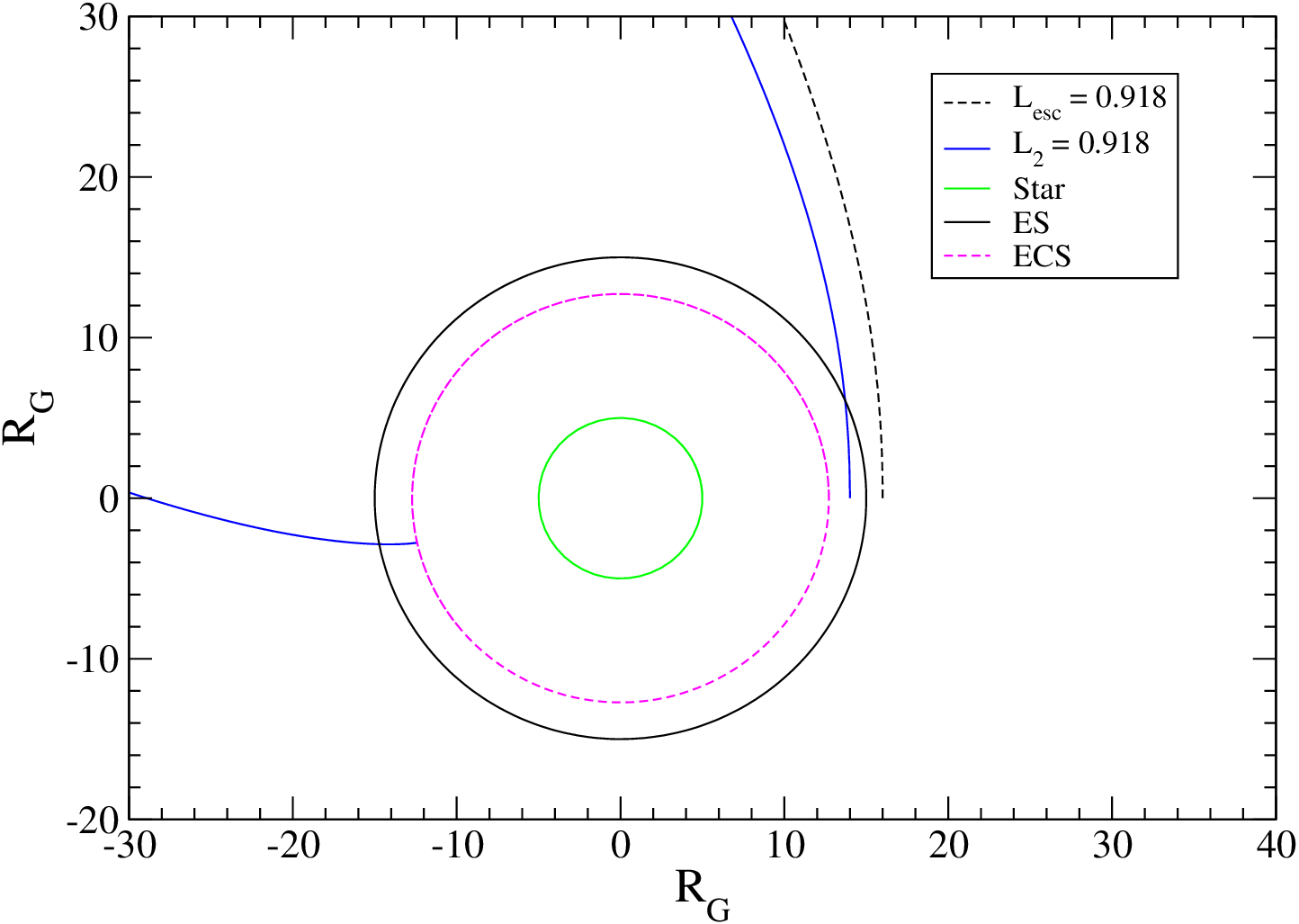}
\caption{Trajectories corresponding to the $L_1 =0.36$ curve in Fig.~\ref{drag} for $L_2 = 0.918$. Black dashed trajectory is unbounded.
The bounded trajectory (blue) ends on an ECS, which in this case is within
the sphere of minimum escape radius (black circle).}
\label{ECS}
\end{figure}

Radiation increases so strongly towards the source in general
relativity that at first sight it seems easy to unbind matter orbiting
a compact luminous star. For instance,
a modest increase in luminosity would be sufficient to expel matter
from the ISCO {\sl if there were no radiation drag}, e.g., a change from 0 to
$0.283\,L_{\rm Edd}$, or from $0.360\,L_{\rm Edd}$ to $0.589\,L_{\rm Edd}$
would suffice, as can be seen in Fig.~\ref{nodrag}.
However, the same radiation very strongly impedes the motion of matter moving
in the optically thin regions illuminated by the star by
exerting a drag that reduces the angular momentum and energy inherent in the motion.
When this effect is properly taken into account, the conclusion is reversed:
it is, in fact, very difficult to expel matter orbiting at the ISCO, typically this is only possible when super-Eddington luminosities are attained. However, the effects of drag fall off very quickly with distance to the radiation source,
so that  an Eddington outburst of a lower luminosity source (say, initial luminosity $<0.3\,L_{\rm Edd}$),  such as an atoll source is sufficient to clear out all test particles orbiting in the optically thin region at $r\gtrsim 10R_G$ (Figs.~\ref{comparison},\ref{drag})

The numerical results obtained in the paper apply to the motion of test particles. However, they should be also applicable, at least qualitatively, to optically thin plasma, e.g., the  coronae of accretion disks.
The radiation front following an outburst of the source moves at the speed of light appropriate for the medium, which is always larger than both the  sound speed and the speed of the orbiting particles along any trajectory they may follow. All parts of the optically thin plasma will feel the influence of radiation before any significant hydrodynamic interaction occurs between the different regions at various radial distance from the source.

First let us consider plasma outside the spherical surface of minimum escape radius.
Assuming that the plasma is distributed axisymmetrically, we can rule out intersection of outgoing trajectories from a ring of plasma originally located at a given radial distance from the source. Further, as the effects of drag fall off with distance, the motion of the  outer rings of plasma will not be impeded by that of the inner rings. However, it is true that the inner rings will at first travel faster than the outer rings, because the outward velocity shortly after the passage of the radiation front scales with initial orbital velocity. This will lead to compressive heating of the plasma, thus robbing it of some kinetic energy and making the escape of the plasma less likely. The expected effect of replacing test particles with plasma is to increase the minimum luminosity change required to make the trajectories unbounded or, equivalently, to increase the minimum escape radius $x_{\rm esc}$ for a given final luminosity.

By the same arguments, plasma will remain bound to the system in those regions from which test particles cannot escape. Thus, we arrive at a robust conclusion that even in Eddington luminosity outbursts radiation drag makes it difficult for matter orbiting near the ISCO to escape the system. However, that region will be temporarily cleared of any optically thin plasma while the plasma follows the initial outwards-going part of its bounded trajectory.

As we have seen in Figs.~\ref{outcome},~\ref{2tracks},~\ref{ECS}, the bounded trajectories can cover a wide range of distance. 
 It seems obvious that under the conditions corresponding to the non-circular bounded trajectories discussed in this paper, colliding shells of plasma will undergo compressive and dissipative heating. The temperatures involved are expected to be quite high, a fraction of the virial energy being involved in the transfer of kinetic to thermal energy \citep{2013A&A...551A..70K}.
Thus, we would like to suggest that strong illumination alone of optically thin plasma may provide a sufficient heating mechanism to explain the origin of the observed high temperatures in the X-ray  emitting coronae of LMXBs.

\section{Conclusions}
\label{conclusion}

We have numerically solved in full GR the equations of motion of a
test particle moving in a strong radiation field.  The motion of test
particles, and most likely also of plasma, around a gravitating
luminous body is strongly affected by rapid changes of its luminosity.
For any initial luminosity and circular orbit, a sufficiently large
increase in luminosity will unbind the particle.  The effects of GR,
and especially of radiation drag, make this ``escape'' luminosity a
strong function of distance for the first several tens of
gravitational radii.  This GR result is in contrast with the Newtonian
case, where the escape luminosity is independent of radius.

In particular, close to the ISCO radiation drag is so strong, that it
takes an about Eddington luminosty outburst to eject particles to infinity.
If the initial luminosity is a sizable fraction of Eddington the final
luminosity required to eject the particle may be super-Eddington.
This may explain why X-ray bursters typically seem to be rebuilding
their inner accretion disks shortly after the outburst.
However, the effects of radiation drag drop very rapidly with distance,
and already at radii $\sim 20R_{G}$ a sub-Eddington burst is typically
sufficient to eject test particles to infinity.

We expect that similar conclusions will also hold for the motion of
plasma, or hot gas, and we intend to verify these expectations
in future work with a hydro code.

\begin{acknowledgements}
    We thank Wenfei Yu for illuminating conversations in the early stages
of this project. Research supported in part by Polish NCN grant
 UMO-34 2011/01/B/ST9/05439.
\end{acknowledgements}
\bibliographystyle{aa}
\bibliography{coronadrag} 
\end{document}